\documentclass[10pt,a4paper]{article}
\usepackage{amsmath}
\usepackage{amssymb}
\usepackage{amsfonts}
\usepackage{amstext}
\usepackage{amsbsy}
\usepackage[mathscr]{eucal}
\usepackage{graphicx}
\usepackage{color}
\usepackage[all]{xy}
\newcommand{\beq}{\begin{equation}}
\newcommand{\eeq}{\end{equation}}
\newcommand{\beqa}{\begin{eqnarray}}
\newcommand{\eeqa}{\end{eqnarray}}
\newcommand{\ket}[1]{| #1 \rangle}

\newtheorem{thm}{Theorem}[subsection]

 \numberwithin{equation}{subsection}

\makeindex
\title{\Large\textbf{Quantum gate entangler for general  multipartite systems}}

\author{\textit{ Hoshang Heydari}\\
        \small\textit{Institute of Quantum
Science, Nihon University,}\\
\small\textit{1-8 Kanda-Surugadai, Chiyoda-ku, Tokyo 101-8308, Japan
}\\\small\textit{Email: hoshang@edu.cst.nihon-u.ac.jp}}

\date{}

\begin{document}

\maketitle \thispagestyle{empty}

\begin{abstract}
We construct quantum gate entangler for general multipartite states based on topological unitary operators. We show that these operators can entangle quantum states if they satisfy the separability condition that is given by the complex multi-projective Segre variety. We also in detail discuss the construction of quantum gate entangler for higher dimensional bipartite and three-partite quantum systems.
\end{abstract}

\section{Introduction}
In the field of quantum computing and quantum information entangled states are used as resource to accomplish interesting task such as quantum teleportation. Entangled states can also be used as building block of quantum computer. The quantum gates entangler  will have a central  role in construction of  such entangled based scheme for quantum computer. Topological quantum computer also has been considered by many researcher and there are several  proposals for  construction of such a fault-tolerant quantum computer. Recently, L. Kauffman and S. Lomonaco Jr. have constructed topological quantum gate entangler for two-qubit state \cite{kauf}. These  topological operator are called
braiding operator that can entangle quantum state. These operator
are also unitary solution of Yang-Baxter equation. Thus these topological unitary transformations  are very suitable for
application in the field of quantum computing.  We have also
recently construct such braiding operators for multi-qubit states \cite{Hosh7}. In this paper, we will construct quantum entangler for general multipartite states based on these braiding operators. In particular in section \ref{com} we will give short introduction to the basic definition of complex projective space and complex projective variety. We also review the construction of complex multi-projective Segre embedding and Segre variety. In section \ref{ent} we will give an introduction to the construction of Artin braid group and Yang-Baxter equation. The section \ref{gen} is the main part of this paper where we will construct topological unitary operators that can entangle a general multipartite quantum state. Finally, in section \ref{exa} we visualize our construction with some examples of construction  of these operators for higher dimensional bipartite and multipartite states.

\section{Complex projective space and Segre variety}\label{com}
In this section, we will introduced the basic definition of complex projective space,
variety, and ideal.
Here are some prerequisites on projective algebraic geometry
\cite{Griff78,Mum76}. We will also review the construction of the
Segre variety for
general pure multipartite states \cite{Hosh5,Hosh6}.
Let $C[z]=C[z_{1},z_{2}, \ldots,z_{n}]$ denotes the polynomial
algebra in $n$  variables with complex coefficients. Then, given a
set of $q$ polynomials $\{g_{1},g_{2},\ldots,g_{q}\}$ with $g_{i}\in
C[z]$, we define a complex affine variety as
\begin{eqnarray}
&&\mathcal{V}_{\mathbf{C}}(g_{1},g_{2},\ldots,g_{q})=\{P\in\mathbb{C}^{n}:
g_{i}(P)=0~\forall~1\leq i\leq q\},
\end{eqnarray}
where $P=(a_{1},a_{2}, \ldots,a_{n})$ is called a point of
$\mathbb{C}^{n}$ and the $a_{i}$ are called the coordinates of $P$.
A complex projective space $\mathbb{P}_{\mathbb{C}}^{n}$ is defined
to be the set of lines through the origin in $\mathbb{C}^{n+1}$,
that is,
$
\mathbb{P}_{\mathbb{C}}^{n}=\mathbb{C}^{n+1}-{0}/\sim$, where the equivalence relation $\sim$ is defined as
$(x_{1},\ldots,x_{n+1})\sim(y_{1},\ldots,y_{n+1},~\lambda\in
\mathbb{C}-0,~y_{i}=\lambda x_{i} ~\forall ~0\leq i\leq n+1.
$.
Given a set of homogeneous polynomials
$\{g_{1},g_{2},\ldots,g_{q}\}$  with $g_{i}\in C[z]$, we define a
complex projective variety as
\begin{eqnarray}
&&\mathcal{V}(g_{1},\ldots,g_{q})=\{O\in\mathbb{P}_{\mathbb{C}}^{n}:
g_{i}(O)=0~\forall~1\leq i\leq q\},
\end{eqnarray}
where $O=[a_{1},a_{2},\ldots,a_{n+1}]$ denotes the equivalent class
of point $\{\alpha_{1},\alpha_{2},\ldots,$
$\alpha_{n+1}\}\in\mathbb{C}^{n+1}$. We can view the affine complex
variety
$\mathcal{V}_{\mathbb{C}}(g_{1},g_{2},\ldots,g_{q})\subset\mathbb{C}^{n+1}$
as a complex cone over the complex projective variety
$\mathcal{V}(g_{1},g_{2},\ldots,g_{q})$.


 The Segre map is give by
\begin{equation}
\begin{array}{ccc}
  \mathcal{S}_{N_{1},\ldots,N_{m}}:\mathbb{P}_{\mathbb{C}}^{N_{1}-1}\times\mathbb{P}_{\mathbb{C}}^{N_{2}-1}
\times\cdots\times\mathbb{P}_{\mathbb{C}}^{N_{m}-1}&\longrightarrow&
\mathbb{P}_{\mathbb{C}}^{N_{1}N_{2}\cdots N_{m}-1}\\
 ((\alpha^{1}_{1},\alpha^{1}_{2},\ldots,\alpha^{1}_{N_{1}}),\ldots,
 (\alpha^{m}_{1},\alpha^{m}_{2},\ldots,\alpha^{m}_{N_{m}})) & \longmapsto&
 (\alpha^{1}_{i_{1}}\alpha^{2}_{i_{2}}\cdots\alpha^{m}_{i_{m}}). \\
\end{array}
\end{equation}
Now, let $\alpha_{i_{1}i_{2}\cdots i_{m}}$,$1\leq i_{j}\leq N_{j}$
be a homogeneous coordinate-function on
$\mathbb{P}_{\mathbb{C}}^{N_{1}N_{2}\cdots N_{m}-1}$.
Moreover, denote a general, composite quantum system with $m$ subsystems as
$\mathcal{Q}=\mathcal{Q}^{p}_{m}(N_{1},N_{2},\ldots,N_{m})
=\mathcal{Q}_{1}\mathcal{Q}_{2}\cdots\mathcal{Q}_{m}$, with the pure
state $ \ket{\Psi}=\sum^{N_{1}}_{k_{1}=1}$ $\sum^{N_{2}}_{k_{2}=1} $
$\cdots\sum^{N_{m}}_{k_{m}=1}\alpha_{k_{1}k_{2}\ldots k_{m}}
\ket{k_{1}}\otimes\ket{k_{2}}\otimes\cdots\otimes \ket{k_{m}} $ and
corresponding to the Hilbert space $
\mathcal{H}_{\mathcal{Q}}=\mathcal{H}_{\mathcal{Q}_{1}}\otimes
\mathcal{H}_{\mathcal{Q}_{2}}\otimes\cdots\otimes\mathcal{H}_{\mathcal{Q}_{m}}
$, where the dimension of the $j$th Hilbert space is
$N_{j}=\dim(\mathcal{H}_{\mathcal{Q}_{j}})$. Furthermore, let
$\rho_{\mathcal{Q}}$ denote a density operator acting on
$\mathcal{H}_{\mathcal{Q}}$. The density operator
$\rho_{\mathcal{Q}}$ is said to be fully separable, which we will
denote by $\rho^{sep}_{\mathcal{Q}}$, with respect to the Hilbert
space decomposition, if it can  be written as $
\rho^{sep}_{\mathcal{Q}}=\sum^\mathrm{N}_{k=1}p_k
\bigotimes^m_{j=1}\rho^k_{\mathcal{Q}_{j}},~\sum^\mathrm{N}_{k=1}p_{k}=1
$
 for some positive integer $\mathrm{N}$, where $p_{k}$ are positive real
numbers and $\rho^k_{\mathcal{Q}_{j}}$ denotes a density operator on
Hilbert space $\mathcal{H}_{\mathcal{Q}_{j}}$. If
$\rho^{p}_{\mathcal{Q}}$ represents a pure state, then the quantum
system is fully separable if $\rho^{p}_{\mathcal{Q}}$ can be written
as
$\rho^{sep}_{\mathcal{Q}}=\bigotimes^m_{j=1}\rho_{\mathcal{Q}_{j}}$,
where $\rho_{\mathcal{Q}_{j}}$ is the density operator on
$\mathcal{H}_{\mathcal{Q}_{j}}$. If a state is not separable, then
it is said to be an entangled state.
Now, let
us consider the composite quantum system
$\mathcal{Q}^{p}_{m}$ $(N_{1},N_{2},\ldots,N_{m})$ and let $
\mathcal{A}=\left(\alpha_{i_{1}i_{2}\ldots i_{m}}\right)_{1\leq
i_{j}\leq N_{j}}, $ for all $j=1,2,\ldots,m$. $\mathcal{A}$ can be
realized as the following set $\{(i_{1},i_{2},\ldots,i_{m}):1\leq
i_{j}\leq N_{j},\forall~j\}$, in which each point
$(i_{1},i_{2},\ldots,i_{m})$ is assigned the value
$\alpha_{i_{1}i_{2}\ldots i_{m}}$. This realization of $\mathcal{A}$
is called an $m$-dimensional box-shape matrix of size $N_{1}\times
N_{2}\times\cdots\times N_{m}$, where we associate to each such
matrix a sub-ring
$\mathrm{S}_{\mathcal{A}}=\mathbb{C}[\mathcal{A}]\subset\mathrm{S}$,
where $\mathrm{S}$ is a commutative ring over the complex number
field. Then the ideal $\mathcal{I}^{m}_{\mathcal{A}}$ of
$\mathrm{S}_{\mathcal{A}}$ is generated by
\begin{equation}
\mathcal{I}^{m}_{\mathcal{A}}=
\left\langle\alpha_{k_{1}k_{2}\ldots k_{m}}\alpha_{l_{1}l_{2}\ldots l_{m}}-
\alpha_{k_{1}k_{2}\ldots k_{j-1}l_{j}k_{j+1}\ldots
k_{m}}\alpha_{l_{1}l_{2} \ldots l_{j-1} k_{j}l_{j+1}\ldots l_{m}}\right\rangle.
\end{equation}
The ideal $\mathcal{I}^{m}_{\mathcal{A}}$
describes the separable states in
$\mathbb{P}_{\mathbb{C}}^{N_{1}N_{2}\cdots N_{m}-1}$ \cite{Hosh5,Hosh6}.
The image of the Segre embedding
$\mathrm{Im}(\mathcal{S}_{N_{1},N_{2},\ldots,N_{m}})$, which again
is an intersection of families of quadric hypersurfaces in
$\mathbb{P}_{\mathbb{C}}^{N_{1}N_{2}\cdots N_{m}-1}$, is called
Segre variety and it is given by
\begin{equation}\label{eq: submeasure}
\mathrm{Im}(\mathcal{S}_{N_{1},N_{2},\ldots,N_{m}})=\bigcap_{\forall
j}\mathcal{V}\left(\alpha_{k_{1}k_{2}\ldots k_{m}}\alpha_{l_{1}l_{2}\ldots l_{m}}-
\alpha_{k_{1}k_{2}\ldots l_{j}\ldots
k_{m}}\alpha_{l_{1}l_{2} \ldots  k_{j}\ldots l_{m}}\right).
\end{equation}
This result is very important for proof of main theorem of this paper, namely the construction
of general quantum gate entangler for general multipartite states.

\section{Artin braiding operator and Yang-Baxter equation}\label{ent}
In this section we will give a short introduction to Artin braid
group and Yang-Baxter equation. We will study relation between
topological and quantum entanglement by investigating the unitary
representation of Artin braid group. Here are some general
references on quantum group and low-dimensional topology
\cite{kassel,chari}. The Artin braid group $\mathrm{B}_{n}$ on $n$
strands is generated by $\{b_{n}:1\leq i\leq n-1\}$ and we have
the following relations in the group $\mathrm{B}_{n}$:
 \begin{description}
   \item[i)] $
    b_{i}b_{j}=b_{j}b_{i}$ for $|i-j|\geq n $
and
   \item[ii)]  $b_{i}b_{i+1}b_{i}=b_{i+1}b_{i}b_{i+1}$ for $ 1\leq i<n$.
 \end{description}
 Let $\mathcal{V}$ be a complex vector space.
  Then, for two
strand braid there is associated an operator
$\mathcal{R}:\mathcal{V}\otimes\mathcal{V}\longrightarrow\mathcal{V}\otimes\mathcal{V}$.
Moreover, let $\mathcal{I}$ be the identity operator on
$\mathcal{V}$. Then, the Yang-Baxter equation is defined by
\begin{equation}\label{YB}
    (\mathcal{R}\otimes \mathcal{I})(\mathcal{I}\otimes \mathcal{R})(\mathcal{R}\otimes
    \mathcal{I})=(\mathcal{I}\otimes \mathcal{R})(\mathcal{R}\otimes \mathcal{I})(\mathcal{I}\otimes
    \mathcal{R}).
\end{equation}
The Yang-Baxter equation represents the fundamental topological
relation in the Artin braid group. The inverse to $\mathcal{R}$ will
be associated with the reverse elementary braid on two strands.
Next, we define a representation $\tau$ of the Artin braid group to
the automorphism of $\mathcal{V}^{\otimes
m}=\mathcal{V}\otimes\mathcal{V}\otimes\cdots\otimes\mathcal{V}$ by
\begin{equation}\label{ABG}
    \tau(b_{i})=\mathcal{I}\otimes\cdots\otimes\mathcal{I}\otimes\mathcal{R}\otimes\mathcal{I}\otimes\cdots\otimes\mathcal{I},
\end{equation}
where $\mathcal{R}$ are in position $i$ and $i+1$. This equation
describe a representation of the braid group if $\mathcal{R}$
satisfies the Yang-Baxter equation and is also invertible.
Moreover, this representation of braid group is unitary if
$\mathcal{R}$ is also unitary operator. Thus $\mathcal{R}$ being
unitary indicated that this operator can performs topological
entanglement and it also can be considers as quantum gate. It has
been show in \cite{kauf} that $\mathcal{R}$ can also perform
quantum entanglement by acting on qubits states.
Note also that  $\mathcal{R}$ is a solution to
the braided version of the Yang-Baxter equation and $\tau=\mathcal{P}\mathcal{R}$ is a
solution to the algebraic Yang-Baxter equation and $\mathcal{P}$
represent a virtual or flat crossing.
In general, let $M=(\mathcal{M}_{kl})$ denote an $n\times n$
matrix with complex elements and let $\mathcal{R}$ be defined by
$\mathcal{R}^{kl}_{rs}=\delta^{k}_{s}\delta^{l}_{r}\mathcal{M}_{kl}$.
Then $\mathcal{R}$ is a unitary solution to the Yang-Baxter
equation.

\section{General multipartite quantum gate entangler}\label{gen}
In this section we will construct a topological unitary operator that entangle a general multipartite state. This an extension of our recent work on construction of such a unitary transformation for multi-qubit state.  Moreover, we can only construct this operator for quantum system with the same dimension. That is, let us consider quantum system $\mathcal{Q}^{p}_{m}(N,N,\ldots,N)$, where $N_{1}=N_{2}=\cdots=N_{m}=N$. Then,
for a general multipartite state a topological unitary
transformation $\mathcal{R}_{N^{m}\times N^{m}}$ that create
multipartite entangled state is defined by
$\mathcal{R}_{N^{m}\times N^{m}}=\mathcal{R}^{d}_{N^{m}\times N^{m}}+\mathcal{R}^{ad}_{\mathcal{N}\times \mathcal{N}}$, where
\begin{equation}
\mathcal{R}^{a}_{N^{m}\times N^{m}}=\mathrm{diag}(\alpha_{11\cdots1},0,\ldots,0,\alpha_{NN\cdots N})
\end{equation}
 is a
diagonal matrix and
\begin{equation}\mathcal{R}^{ad}_{N^{m}\times N^{m}}=\mathrm{antidiag}(0,\alpha_{NN\cdots N-1},\alpha_{NN\cdots N-2},
\ldots,\alpha_{1\cdots1 3},\alpha_{1\cdots1 2},0)
\end{equation}
 is an anti-diagonal matrix.
 Then we have following theorem for general multipartite
states.
\begin{thm} If elements of $\mathcal{R}_{N^{m}\times N^{m}}$ satisfies
\begin{eqnarray}\label{segreply1}
&&
\alpha_{k_{1}k_{2}\ldots k_{m}}\alpha_{l_{1}l_{2}\ldots l_{m}}\neq
\alpha_{k_{1}k_{2}\ldots k_{j-1}l_{j}k_{j+1}\ldots
k_{m}}\alpha_{l_{1}l_{2} \ldots l_{j-1} k_{j}l_{j+1}\ldots l_{m}},
\end{eqnarray}then the state
$\mathcal{R}_{N^{m}\times N^{m}}(\ket{\psi}\otimes\ket{\psi}\otimes\cdots\otimes\ket{\psi})$, with
$\ket{\psi}=\ket{1}+\ket{2}+\ldots+\ket{N}$ is entangled.
\end{thm}
The proof  follows from the  construction of $\mathcal{R}_{N^{m}\times
N^{m}}$ which is based on completely separable set of
multipartite states defined by the Segre variety
 That is
the state
\begin{equation}
\mathcal{R}_{N^{m}\times
N^{m}}(\ket{\psi}\otimes\ket{\psi}\otimes\cdots\otimes\ket{\psi})= \sum^{N}_{k_{1},k_{2},\ldots,k_{m}=1}\alpha_{k_{1}k_{2}\ldots k_{m}}
\ket{k_{1}}\otimes\ket{k_{2}}\otimes\cdots\otimes \ket{k_{m}}
\end{equation}
is entangled if and only if
\begin{equation}
\alpha_{k_{1}k_{2}\ldots k_{m}}\alpha_{l_{1}l_{2}\ldots l_{m}}-
\alpha_{k_{1}k_{2}\ldots k_{j-1}l_{j}k_{j+1}\ldots
k_{m}}\alpha_{l_{1}l_{2} \ldots l_{j-1} k_{j}l_{j+1}\ldots l_{m}}\neq0
\end{equation}
and this follows directly from the construction of the completely separable state defined by the Segre variety.
 Note that this operator is a quantum gate
entangler since
\begin{eqnarray}
\tau_{N^{m}\times N^{m}}&=&\mathcal{R}_{N^{m}\times
N^{m}}\mathcal{P}_{N^{m}\times N^{m}}\\\nonumber&=&
\mathrm{diag}(\alpha_{1\cdots1 1},\alpha_{1\cdots1 2},\ldots,\alpha_{N\cdots N N})
\end{eqnarray}
 is a $N^{m}\times N^{m}$
phase gate and $\mathcal{P}_{N^{m}\times N^{m}}$
is $N^{m}\times
N^{m}$ a swap gate. This is our main result and in following section
 we will illustrate is by some examples.

\section{Higher dimensional bipartite and three-partite quantum gate entangler}\label{exa}
We will begin with an example of higher dimensional composite bipartite quantum system $\mathcal{Q}^{p}_{m}(3,3)$.
For this bipartite system the unitary operator $\mathcal{R}_{9\times 9}$, is give by
\begin{eqnarray}
\mathcal{R}_{9\times 9}&=&\mathcal{R}^{d}_{9\times 9}+\mathcal{R}^{ad}_{9\times 9}\\\nonumber&=&\mathrm{diag}(\alpha_{11},0,0,0,0,0,0,0,\alpha_{33})\\\nonumber
&&+\mathrm{antidiag}(0,\alpha_{32},\alpha_{31},\alpha_{23},\alpha_{22},\alpha_{21},
\alpha_{13},\alpha_{12},0)\\\nonumber&=&\left(%
\begin{array}{ccccccccc}
  \alpha_{11} & 0 & 0 & 0 & 0 & 0 & 0 & 0&0 \\
  0 & 0 & 0 & 0 & 0 & 0 & 0& \alpha_{32}  &0\\
  0 & 0 & 0 & 0 & 0 & 0& \alpha_{31} & 0 &0\\
  0 & 0 & 0 & 0 & 0& \alpha_{23}  & 0 & 0 &0\\
   0 & 0 & 0 & 0 & \alpha_{22} & 0 & 0 & 0 &0\\
  0 & 0 & 0& \alpha_{21}  & 0 & 0 & 0 & 0 &0\\
   0 & 0& \alpha_{13}  & 0 & 0 & 0 & 0 & 0 &0\\
   0 & \alpha_{12} & 0 & 0 & 0 & 0 & 0 & 0 &0\\
  0 & 0 & 0 & 0 & 0 & 0 & 0 & 0&\alpha_{33}\\
\end{array}%
\right).
\end{eqnarray}
Now, if elements of $\mathcal{R}_{N^{m}\times N^{m}}$ satisfies
$
\alpha_{k_{1}k_{2}}\alpha_{l_{1}l_{2}}\neq
\alpha_{k_{1}l_{2}}\alpha_{l_{1}k_{2}},
$ for $j=1,2$ and $k_{1},k_{2},l_{1},l_{2}=1,2,3$, then the state
\begin{eqnarray}\nonumber
\mathcal{R}_{9 \times 9}(\ket{\psi}\otimes\ket{\psi})&=&\mathcal{R}_{9 \times 9}((\ket{1}+\ket{2}+\ket{3})\otimes(\ket{1}+\ket{2}+\ket{3}))\\&=&
\sum^{3}_{k_{1},k_{2}=1}\alpha_{k_{1}k_{2}}\ket{k_{1}}\otimes \ket{k_{2}}
\end{eqnarray}
 is entangled. But it is the case based on construction of separable
  state for bipartite state defined by the Segre variety. This can be easily construct for any higher dimensional bipartite   states. Note also that the permutation matrix is give by
  \begin{equation}
\mathcal{P}_{9\times 9}=\mathrm{diag}(1,0,0,0,0,0,0,0,1)+\mathrm{antidiag}(0,1,1,1,1,1,1,1,0)
\end{equation}
and the phase gate is give by
$
\tau_{9\times 9}=\mathrm{diag}(\alpha_{11},\alpha_{32},\alpha_{31},\alpha_{23},\alpha_{22},\alpha_{21},
\alpha_{13},\alpha_{12},\alpha_{33})
$
  Next, we will construct a quantum gate entangler for simplest multipartite higher dimensional system, namely the quantum system  $\mathcal{Q}^{p}_{m}(3,3,3)$.
For this three-partite state the unitary operator $\mathcal{R}_{27\times 27}$, is give by
\begin{eqnarray}
\mathcal{R}_{27\times 27}&=&\mathcal{R}^{d}_{27\times 27}+\mathcal{R}^{ad}_{27\times 27}\\\nonumber&=&\mathrm{diag}(\alpha_{111},0,\ldots,0,\alpha_{333})+\mathrm{antidiag}(0,\alpha_{332},\alpha_{331},
\alpha_{323},\alpha_{322},\alpha_{321}\\\nonumber
&&,
\alpha_{313},\alpha_{312},\alpha_{311},\alpha_{233},\alpha_{232},\alpha_{231},
\alpha_{223},\alpha_{222},\alpha_{221},\alpha_{213},\alpha_{212},\alpha_{211}\\\nonumber
&&,
\alpha_{133},\alpha_{132},\alpha_{131},\alpha_{123},\alpha_{122},\alpha_{121}
,\alpha_{113},\alpha_{112},0).
\end{eqnarray}
Now,  the three-partite state
\begin{eqnarray}\nonumber
&&\mathcal{R}_{27\times 27}\left((\ket{1}+\ket{2}+\ket{3})\otimes(\ket{1}+\ket{2}+\ket{3})\otimes(\ket{1}+\ket{2}+\ket{3})\right)\\\nonumber
&&=
\sum^{3}_{k_{1},k_{2},k_{3}=1}\alpha_{k_{1}k_{2}k_{3}}\ket{k_{1}}\otimes \ket{k_{2}}\otimes \ket{k_{3}}
\end{eqnarray}
is entangled if the elements of $\mathcal{R}_{27\times 27}$ satisfy
$
\alpha_{k_{1}k_{2}k_{3}}\alpha_{l_{1}l_{2}l_{3}}\neq
\alpha_{k_{1}l_{j}k_{3}}\alpha_{l_{1}k_{j}l_{3}},
$ for $j=1,2,3$ and $k_{1},k_{2},k_{3},l_{1},l_{2},l_{3}=1,2,3$.
Thus in this paper we have shown that how we can construct quantum gate entangler for general multipartite state based an combination of topological and geometrical method.

\begin{flushleft}
\textbf{Acknowledgments:} The  author gratefully acknowledges the
financial support of the Japan Society for the Promotion of Science
(JSPS).
\end{flushleft}


\end{document}